\documentstyle[latexsym,amssymb,epsfig,prb,eqsecnum,aps]{revtex}

\begin{document}

\def\s {{$\sigma$ phase~}}

\title{Vibrational properties of the one-component $\sigma$ phase}

\author{S.I.Simdyankin$^1$, S.N.Taraskin$^2$, M.Dzugutov$^1$, and
S.R.Elliott$^2$}

\address{
$^1$ Department of Numerical Analysis and Computing Science,\\
     Royal Institute of Technology, SE--100 44 Stockholm, Sweden.\\
$^2$ Department of Chemistry, University of Cambridge,\\
     Lensfield Road, Cambridge CB2 1EW, UK\\
}

\date{\today}

\maketitle

\begin{abstract}
A structural model of a one-component $\sigma$-phase crystal has 
been constructed by means of molecular dynamics simulation. 
The phonon dispersion curves and the vibrational density of 
states were computed for this model. 
The dependence of the vibrational properties on the thermodynamical 
parameters was investigated.
The vibrational density of states of the $\sigma$-phase structure 
is found to be similar to that of a one-component glass 
with icosahedral local order. 
On the basis of this comparison it is concluded that the $\sigma$
phase can be considered to be a good crystalline reference
structure for this glass.
\end{abstract}


\section{Introduction} 
\label{sec:intro}

The local atomic order in disordered condensed materials is well defined 
and governs many physical properties \cite{SRE_book}. 
Quite often, for a disordered marerial,
it is possible to find a corresponding crystal
with similar local and even intermediate-range order 
which give rise to similarities in  
many structural and dynamical features of these two solids.
Such a crystal can be regarded as a reference crystalline structure 
(crystalline counterpart) for the corresponding disordered substance. 
In some cases, the reference structure can be uniquely defined.
The simplest examples are toy structural models with  
force-constant and/or mass disorder. 
In these toy models, the atoms occupy their equilibrium positions 
at the sites of a crystalline lattice (e.g. simple cubic),  
which can be considered to be a reference one 
(see e.g. \cite{Schirmacher98}). 
Another related example is a binary substitutional alloy, 
the reference system for which is a periodic point lattice 
with one of the two atomic species placed at 
the lattice sites \cite{Ehrenreich_76}. 
The disorder in such models does not influence the equilibrium 
positions of the atoms arranged in an ideal crystalline lattice. 
This makes possible the use of approximate analytical 
approaches (e.g. the coherent potential approximation 
\cite{Ehrenreich_76}) to treat the vibrational properties of the models,
provided that the vibrational properties of the counterpart crystal
are known. 

In amorphous solids, or glasses, 
the atoms do not occupy the sites of a crystalline lattice, 
which results in positional disorder.
For these materials, a choice of a reference structure becomes problematic.
Good counterparts can usually be found among the crystalline 
polymorphs having the same (or similar) chemical composition as 
the corresponding glass. 
For example, $\alpha$-cristobalite appears to be a good crystalline 
counterpart for vitreous silica \cite{Ding_98,PRB2_97,Dove_97}. 

The main purpose of this paper is to investigate numerically 
the vibrational properties of a one-component $\sigma$-phase
\cite{FrankKasper58,FrankKasper59,Nelson89} crystal which is
conjectured to be a good crystalline counterpart for a one-component
glass with icosahedral local order (IC glass) 
\cite{Dzugutov92,DzugutovGlass}.

The motivation for this choice of a crystalline counterpart of the IC glass
is the following.
The computational model of the IC glass is based on a simple 
empirical pair interatomic potential \cite{Dzugutov92} 
resembling the effective interionic potentials conjectured for simple 
metallic glass-forming alloys \cite{Hafner87}. 
The use of the same potential allows us to construct models of bcc
and $\sigma$-phase crystals that are stable for a wide range of
thermodynamical parameters \cite{Dzugutov97}.
Of these two crystalline structures, the $\sigma$ phase is expected to be
a good reference structure for the IC glass because 
of the following reasons: 
The supercooled IC liquid (where the interactions between atoms are
described by the same potential \cite{Dzugutov92}) undergoes a
transition either to the IC glass or to a dodecagonal quasicrystal
\cite{Dzugutov93} depending on the quench rate \cite{Dzugutov97}. 
This quasicrystal has similar local structural properties with 
the IC glass \cite{Dzugutov93}.
However, the absence of global periodic order in the quasicrystalline 
phase makes the analysis of its vibrational properties a task of 
comparable complexity to that for the glass itself.
The $\sigma$ phase is one of the closest low-order crystalline
approximants \cite{Janot94} for this dodecagonal quasicrystal 
\cite{RothGaehler98}, which means that these two 
(crystalline and quasicrystalline) structures are built up from the same
structural units.
This implies that the IC glass and the $\sigma$ phase,
being both tetrahedrally closed-packed structures \cite{Nelson89}, 
are nearly isomorphous in terms of local order.

Knowledge of the vibrational properties of the $\sigma$-phase
crystal allows for a direct comparison with those of the IC glass.
The apparent similarity in the vibrational densities of states 
of these two structures gives stronger support for the choice 
of this crystalline counterpart for the IC glass.

The $\sigma$-phase structure, used in our computations, has been
obtained by means of molecular dynamics simulation with the use of an
interatomic pair potential \cite{Dzugutov92}.  
The vibrational properties have been investigated by using both a the
normal-mode analysis and by computing the spectra of appropriate
time-correlation functions.

The paper is arranged as follows: 
The $\sigma$-phase structure is described in Sec.~\ref{S2}.
The model and technical details of the calculations are presented in 
Sec.~\ref{sec:compute}.  
In Sec.~\ref{sec:results} we present the results of the simulations.
Some concluding remarks are contained in Sec.~\ref{sec:conclusions}.
%
%
%
%
%
%
%
%
%
%
%
%

\section{The $\sigma$ phase}
\label{S2}
In this Section, we review the known 
structural and dynamical properties of the $\sigma$ phase. 

%
%
%
%
%
%
%
%
%
%
%
%
\subsection{Structure}
\label{S2a}
%
%
%
%
%

The $\sigma$ phase belongs to an important class of tetrahedrally
close-packed crystallographic structures \cite{Nelson89}, viz.  the
Frank-Kasper phases \cite{FrankKasper58,FrankKasper59}.  
The first coordination shells of the constituent atoms in 
these structures form triangulated (Frank-Kasper) polyhedra 
composed entirely of slightly distorted tetrahedra. 
The four possible coordination numbers ($Z$) in
these structures are $Z=12, 14, 15$ and $16$.  
The least distorted tetrahedra are found in icosahedra 
($Z12$  polyhedra). 
Structures of small clusters of atoms interacting via pairwise 
central potentials favor icosahedral order \cite{Nelson89} 
as having the lowest energy.
The prototype $\sigma$ phase structures are $\beta$-U \cite{Lawson88}
and Cr$_{48}$Fe$_{52}$ \cite{Yakel83}.  
There are $30$ atoms per tetragonal unit cell ($tP30$) with 
$c/a \approx 0.52$, where $c$ and
$a$ are the dimensions of the cell (lattice parameters).  
The space group of this phase is $P4_2/mnm$.  
There are $10$ atoms with the coordination number $12$,  $Z12$
or icosahedra, $16$\  $Z14$ atoms and $4$ \ $Z15$ atoms.  
The $-72^{\circ}$ disclination lines \cite{Nelson89} form a network 
(a major skeleton, in the parlance of Frank and Kasper
\cite{FrankKasper58,FrankKasper59}), where rows of $Z14$ atoms parallel
to the tetragonal $c$-axis thread planar networks of $Z14$ and $Z15$
atoms. 
A projection of the $\sigma$-phase structure down the $c$-axis is
shown in Fig.~\ref{sigmaMD}.

The Frank-Kasper phases share their significant geometrical property of
icosahedral local order with simple metallic glasses 
\cite{Nelson89,Hafner87}. 
Some liquid alloys which form Frank-Kasper phases have a tendency to
freeze into metastable amorphous structures (metallic glasses) 
when quenched sufficiently rapidly \cite{Hafner87}.  
It is now generally well accepted that,
at least in the case of metallic alloys of simple constitution, glass
formation is caused by the incompatibility of local icosahedral
coordination with the translational symmetry in Euclidean space
(geometrical frustration) \cite{Nelson89}. 
There exist statistical
mechanical arguments in favor of this scenario of glass formation
\cite{Sachdev85} based on a Landau free-energy analysis.  
The average coordination number, $\overline{Z}$, 
in the Frank-Kasper phases ($13.333 \le \overline{Z} \le 13.5$) 
is very close to that of a sphere-packed
``ideal glass'' model \cite{Nelson83} 
($\overline{Z}_{{\rm ideal}} = 13.4$).  
In a sense, this ``ideal glass'' could be regarded as a
Frank-Kasper phase with an infinitely large unit cell
\cite{Nelson89,Nelson83}. Thus the class of Frank-Kasper 
phases is a natural choice for reference crystalline structures for 
metallic glasses of simple constitution. 

From a structural point of view, the $\sigma$ phase
can be also regarded as a crystalline low-order approximant for dodecagonal
quasicrystals \cite{RothGaehler98}. 
Such quasicrystals \cite{Janot94}, morphologically close to Frank-Kasper
phases, represent an alternative class of noncrystallographic
structures which combine icosahedral local order with 
non-translational long-range order manifested by infinitely 
sharp diffraction peaks. 

\subsection{Dynamics}
\label{S2b}
%
%
%
%
%


Similarities in the local structure of metallic glasses and
Frank-Kasper phases are reflected in the dynamical properties of these
materials. 

The available data about the vibrational dynamics of Frank-Kasper
phases is limited to some of the Laves phases,
a subclass of the Frank-Kasper phases \cite{Hafner87,Kress87}.
For instance, the similarity between the phonon-dispersion
relations of the Mg$_{70}$Zn$_{30}$ glass \cite{SuckRudin83,Suck92}
and those of the Laves phase MgZn$_2$ \cite{Eschrig72} was
emphasized by Hafner \cite{Hafner87,Hafner83}.

Another interesting aspect of the dynamics of Frank-Kasper phases 
is related to the appearance of soft vibrational modes 
in these materials.  
Such a soft low-frequency optic mode at the $\Gamma$-point 
(the origin of the reciprocal lattice) 
has been found numerically in the same Laves phase MgZn$_2$ \cite{Eschrig72}. 
The frequency of this mode decreases with increasing pressure 
(accompanied by volume compression)
and eventually becomes negative, indicating 
a structural phase transition 
\cite{Dove93,Ashcroft76,Heiming91,Zhang95,Meyer98}.  
The authors of Ref.~\cite{Eschrig72} applied 
a group-theoretical analysis and demonstrated 
that the polarization vector of the soft optic mode in MgZn$_2$ is 
determined by the structure symmetry and is independent of interatomic
interactions.  
This suggests that the soft-mode character of some vibrations 
is a generic property of Frank-Kasper phases. 
However, no soft-mode behavior was observed in the
isostructural CaMg$_2$ \cite{Eschrig77} where the mass ratio of 
the constituent elements is reversed with respect to MgZn$_2$. 
Hafner \cite{Hafner87}
suggested that the soft modes in MgZn$_2$ should be attributed to the
relatively large mass of the Zn atoms. This is an example 
of the chemical composition of the materials 
introducing considerable difficulties into the analysis of 
the interplay of the structure and the dynamics. 

A numerical simulation of a one-component Frank-Kasper phase allows us
to eliminate this uncertainty. We investigate the behavior of the
lowest-frequency optic modes in the one-component $\sigma$ phase with
variable pressure in Sec.~\ref{subsubsec:softModes}.

%
%
%
%
%
%
%
%
\section{Methods of computation}
\label{sec:compute}
We have constructed a thermodynamically stable structural model of the
one-component $\sigma$ phase by means of classical molecular
dynamics simulation.
In this case, success strongly depends on the choice of the 
interatomic potential for the model. 
For example, the Lennard-Jones potential, widely used in creating
simple models of liquids \cite{Hansen86}, 
glasses \cite{Sampoli98}, and crystals \cite{Klein78}, is
not suitable for this purpose, because the $\sigma$ phase is not
stable for it in the range of thermodynamical parameters investigated
below (the stable phase is the fcc lattice).
Instead, we use a pair interatomic potential suggested in
Ref.~\cite{Dzugutov92} and show that it is possible to construct a
one-component $\sigma$ phase which is stable for a wide range
of thermodynamical parameters \cite{Dzugutov97}.

\subsection{Model}
\label{sec:model}

As a mathematical model for the study of atomic dynamics in the
crystalline $\sigma$-phase structure, we consider a classical system of
$N$ identical particles interacting via a spherically symmetric pair
potential.

\begin{figure}[b!]
\centerline{\epsfig{file=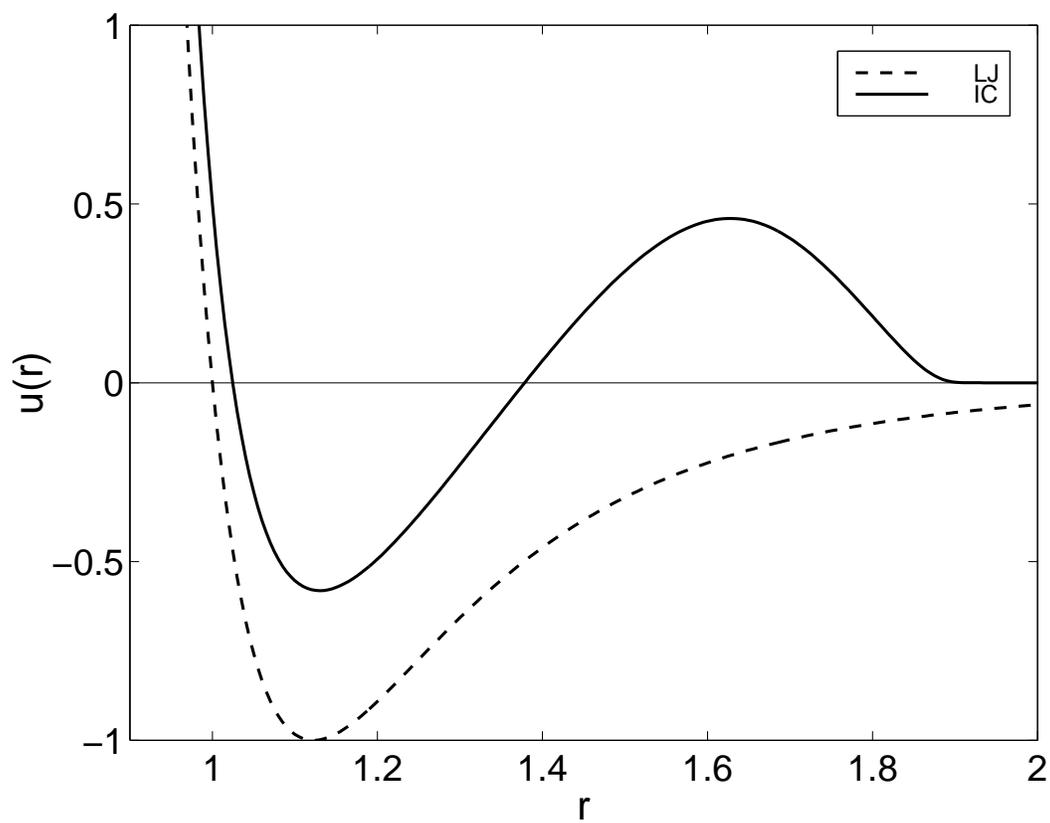,width=14cm,clip=}}
\vskip 20pt
\caption[]{The IC pair potential used in this study
           \protect\cite{Dzugutov92} (solid line) compared with the
           Lennard-Jones potential (dashed line).  }
\label{IC} 
\end{figure}

The pair potential used in this study \cite{Dzugutov92} is designed to
favor icosahedral local order.  The main repulsive part of this
potential is identical to that of the Lennard-Jones potential
$u_{{\rm LJ}}(r) = 4\epsilon[(\sigma/r)^{12}-(\sigma/r)^{6}]$;
therefore all the quantities in these simulations are expressed in 
reduced Lennard-Jones units \cite{Allen87}, i.e. with $\sigma$,
$\epsilon$, and $\tau_0 = (m\sigma^2/\epsilon)^{1/2}$ chosen as
length, energy, and time units.  To convert the reduced units to
physical units one can refer to argon ($m$ = 39.948 a.m.u.)  by
choosing the Lennard-Jones parameters $\sigma=3.4$~\AA~and
$\epsilon/k_B=120$~K.  In this case, our frequency unit
$\nu_0=\tau_0^{-1}$ corresponds to 0.4648~THz.  
The analytical expression defining the potential
is given in Ref.~\cite{Dzugutov92}.
This potential resembles those for simple glass-forming metallic
alloys \cite{Hafner87} with only the first of the Friedel oscillations
being retained (see Fig.~\ref{IC}).

We use conventional molecular dynamics simulations \cite{Allen87} in
which the Newtonian equations of motion are solved using a
finite-difference algorithm with time step equal to 0.01 while
the particles are enclosed in a simulation box of volume $V$ with
periodic boundary conditions. In this case, the total energy $E$ is a
constant of motion and time averages obtained in the course of
simulations approximate the ensemble averages in the microcanonical
(constant-$NVE$) statistical ensemble.  Wavevectors, 
\begin{equation}
{\bf Q} = n_x {\bf Q}_{x,0} + n_y {\bf Q}_{y,0} + n_z {\bf Q}_{z,0}
\label{Qdef}
\end{equation}
with $n_x$, $n_y$ and $n_z$ being integers,
compatible with periodic boundary conditions are multiples of the three 
fundamental wavevectors 
${\bf Q}_{x,0} = \frac{2\pi}{L_x}(1,0,0)$, 
${\bf Q}_{y,0} = \frac{2\pi}{L_y}(0,1,0)$ and
${\bf Q}_{z,0} = \frac{2\pi}{L_z}(0,0,1)$, 
where $L_x$, $L_y$, and $L_z$ are the
dimensions of the (tetragonal) simulation box.  In order to have
a sufficient number of allowed wavevectors within the Brillouin zone, the
sample dimensions must be sufficiently large. The time-correlation 
functions resulted from the molecular
dynamics simulation reported below were obtained for a system of 20580
particles ($7\times7\times14$ unit cells with $30$ atoms per unit cell). 
This relatively large system size also gives sufficient statistical 
accuracy \cite{Dzugutov91}. 
Where necessary, we performed simulations in other ensembles by
modifying the equations of motion \cite{Frenkel96,Smith96}.

The arrangement of atoms in a unit cell of the model $\sigma$-phase
structure used in our computer simulations is shown in
fig.~\ref{sigmaMD}(b).  
The optimal (with respect to an energy minimization) 
$c/a$ ratio was taken to be equal to 0.5273. 
To determine the optimal model structure, it was sufficient to perform
molecular dynamics simulations with the number of particles 
in the system equal to $N=1620$ ($3 \times 3\times 6$ unit cells).
Details of the preparation of this atomic configuration 
are given in section \ref{sec:prepare}.

\subsection{Time-correlation functions}
\label{sec:MD}

A straightforward method to analyze the vibrational dynamics in a
molecular dynamics model is to imitate inelastic neutron scattering
experiments by calculating the dynamical structure factor
$S({\bf Q},\omega)$ \cite{Klein78}, proportional to the neutron
scattering cross-section \cite{Ashcroft76,Hansen86}, which is the
spectrum of the density autocorrelation function:
\begin{equation}
F({\bf Q},t) = <\rho({\bf Q},t)\rho(-{\bf Q},0)> 
\label{F(Q,t)}
\end{equation}
where 
\begin{equation}
\rho({\bf Q},t)=\sum_{k=1}^{N}\exp(-i{\bf Q}\cdot{\bf r}_k(t))
\label{rho(Q,t)}
\end{equation}
is the Fourier transform of the local particle density
\cite{Hansen86}, $N$ is the number of particles in the system,
${\bf r}_k(t)$ is the position vector of particle $k$,  
and the wavevector ${\bf Q}$ takes on the values according to 
Eq.~(\ref{Qdef}).
A longitudinal phonon is associated with a maximum in
$S({\bf Q},\omega)$ at a fixed ${\bf Q}$.  In order to get
information about the transverse modes from $S({\bf Q},\omega)$,
one has to select wavevectors outside the first Brillouin zone
\cite{Schober93}.

In a more convenient way, the vibrational modes can be studied using
the current autocorrelation function \cite{Hansen86}:
\begin{equation}
C_{{\bf e}}({\bf Q},t) = \frac{Q^2}{N} 
<j_{{\bf e}}({\bf Q},t)j_{{\bf e}}(-{\bf Q},0)>
\label{C(Q,t)}
\end{equation}
where
\begin{equation}
j_{{\bf e}}({\bf Q},t) = 
\sum_{k=1}^{N} ({\bf e}\cdot{\bf v}_k(t))
\exp[-i{\bf Q} \cdot {\bf r}_k(t)]
\label{j(Q,t)}
\end{equation}
is the Fourier transform of the local current, ${\bf v}_k(t)$ is
the velocity of particle $k$, and ${\bf e}$ is the unit polarization
vector.  

Note that for the longitudinal polarization,
${\bf e}\parallel{\bf Q}$, Eq.~(\ref{C(Q,t)}) can be obtained from
(\ref{F(Q,t)}) by double differentiation with respect to time.  For
the transverse-current correlation functions, the polarization vector
must be chosen consistently with the lattice symmetry.

At a temperature $T$, the vibrational density of states $g(\omega)$
can be calculated as the Fourier transform of the normalized velocity
autocorrelation function \cite{Dove93,Sampoli98}:
\begin{equation}
Z(t) = \frac{1}{3NT}\left< \sum_{k=1}^{N}
{\bf v}_k(t)\cdot{\bf v}_k(0) \right>
\label{Z(t)}
\end{equation}

We computed the time-correlation functions using the overlapped data
collection technique \cite{Rapaport95}.  The number of overlapped
measurements used for statistical averaging was about 10000. The time
origins of the measurements were separated by 0.2 r.u.  (20 time
steps).

In order to reduce the finite-time truncation effects in the spectra
of the time-correlation functions, we used a Gaussian window function
with the half-width equal to 3 r.u.

\subsection{Normal-mode analysis}
\label{sec:NMA}

To calculate the dispersion relations in the harmonic approximation, we
used the standard method based on diagonalization of the Fourier
transformed dynamical matrix \cite{Maradudin71}.
From the known dispersion relations, $\omega_{s}({\bf Q})$, $s =
1,2,\dots,90$, the vibrational density of states can be computed by
integration over the first Brillouin zone according to
\begin{equation}
g(\omega) = \frac{v}{r(2\pi)^3}\sum_{s=1}^{r}
\int_{{\rm BZ}} \delta(\omega - \omega_{s}({\bf Q})) \; d{\bf Q}
\label{gomega}
\end{equation}
where the sum is over all $r$ dispersion branches 
and $v$ stands for the volume of the unit cell. 
In the computations, a Gaussian function with small but finite width
is substituted for the $\delta$-function.  The half-width of the
Gaussian function in our computations was equal to about 0.05.
Formally, the wavevector ${\bf Q}$ in Eq.~(\ref{gomega}) is a
continuous variable, but in the simulations the integral was estimated
by a sum over a uniform rectangular grid of $100\times100\times100$
points in the first Brillouin zone.

%
%
%
%
%
\section{Results}
\label{sec:results}
\subsection{Optimization of the structure in the $\sigma$ phase}
\label{sec:prepare}

In this subsection, we describe the method of construction of the
thermodynamically stable model of the $\sigma$ phase and analyse the
range of its stability.

\begin{figure}[b!]
\centerline{\epsfig{file=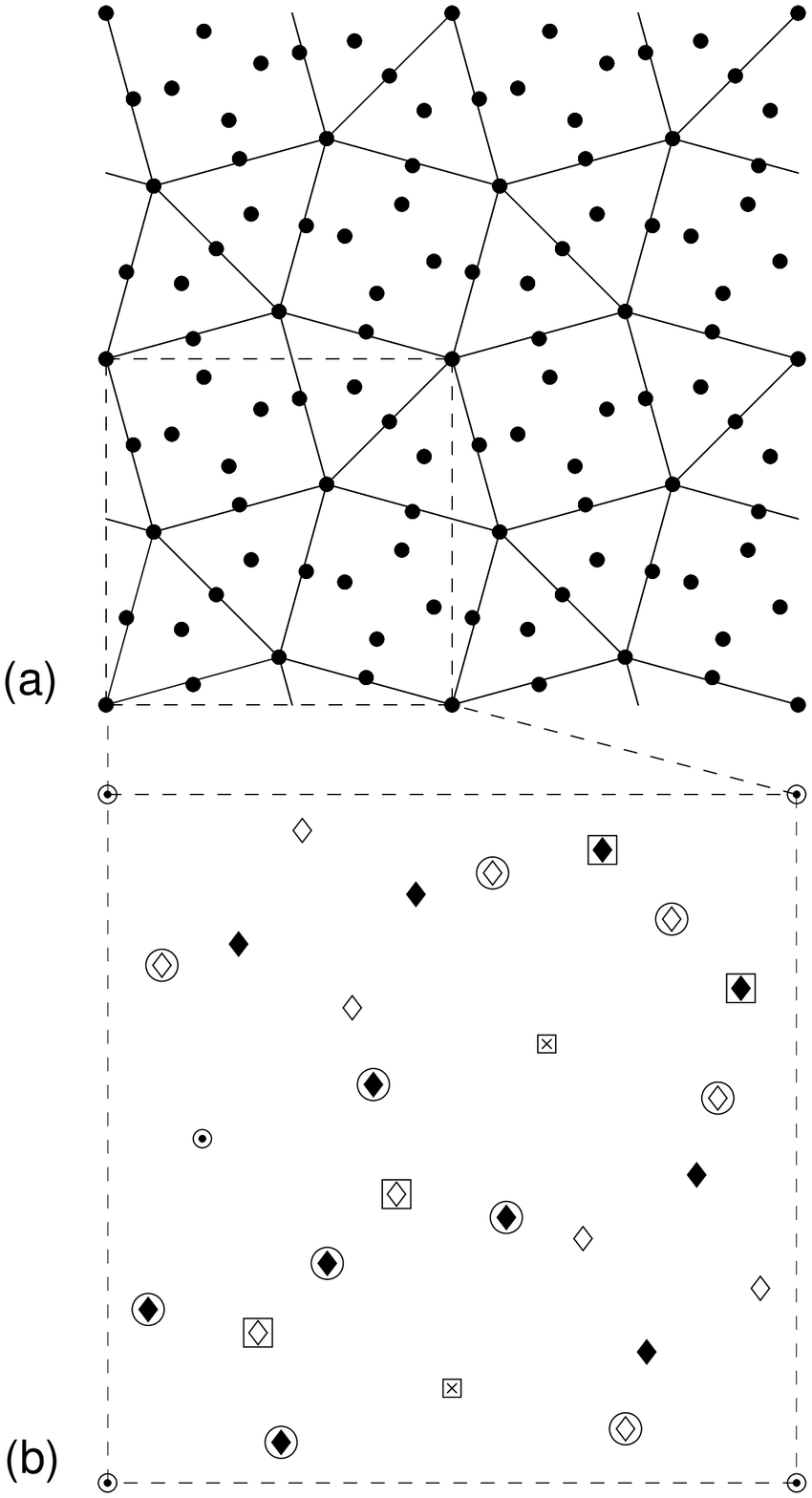,height=12cm}}
\vskip 20pt
\caption[]{Projection down the $c$-axis of the $\sigma$-phase structure:
           (a) $3^2,4,3,4$ net (the numerical symbols are Schl\"afli
           symbols \protect\cite{FrankKasper58}, specifying the number and
           sequence of various polygons around each vertex). The dashed
           square outlines a unit cell; (b) Atomic arrangement in one
           unit (cubic) cell.  $\lozenge$: $z=0$, $\blacklozenge$:
           $z=0.5$, {\footnotesize $\bullet$}: $z=0.2499$. $\times$:
           $z=0.2501$, {\Large $\circ$}: $z=0.7499$, $\Box$:
           $z=0.7501$, $\bigcirc$: Z12 atoms, {\large $\square$}: Z15
           atoms. The rest of atoms are Z14.  Multiplying $z$ by the
           proper $c/a$ ratio gives a tetragonal unit cell with $a=1$.}
\label{sigmaMD} 
\end{figure}

There are several ways to obtain numerical values of the atomic
coordinates in the $\sigma$ phase.  
One way is to use those available for Cr$_{48}$Fe$_{52}$ \cite{DaamsAtlas}. 
Alternatively, a unit cell of the $\sigma$-phase structure can be 
constructed either by manipulating the kagom\'e tiling according to 
the algorithm given by Frank and Kasper in ref.~\cite{FrankKasper59} 
or by stacking the square and triangular basic elements of the dodecagonal 
quasicrystal model \cite{RothGaehler98,Gaehler} into the $3^2,4,3,4$
square-triangle net \cite{FrankKasper59} (see fig.~\ref{sigmaMD}(a)). 
The arrangements of the atoms resulting from these constructions do not 
correspond exactly, although the difference is rather small -- 
the root-mean-square distance between the corresponding atoms in 
different configurations is of the order of a few percent of the  
root-mean-square distance between different atoms in the same configuration.  
In either case, the resulting structure is an approximate one in the
sense that the atomic positions do not correspond to a minimal
potential-energy configuration for a given interaction potential.
To obtain the true structure corresponding to the potential, the
approximate configuration must be relaxed by a molecular dynamics
program.
Moreover the $c/a$ ratio is slightly different for different natural
$\sigma$-phase crystals, which means that this ratio is not uniquely
defined.

\begin{figure}[b!]
\centerline{\epsfig{file=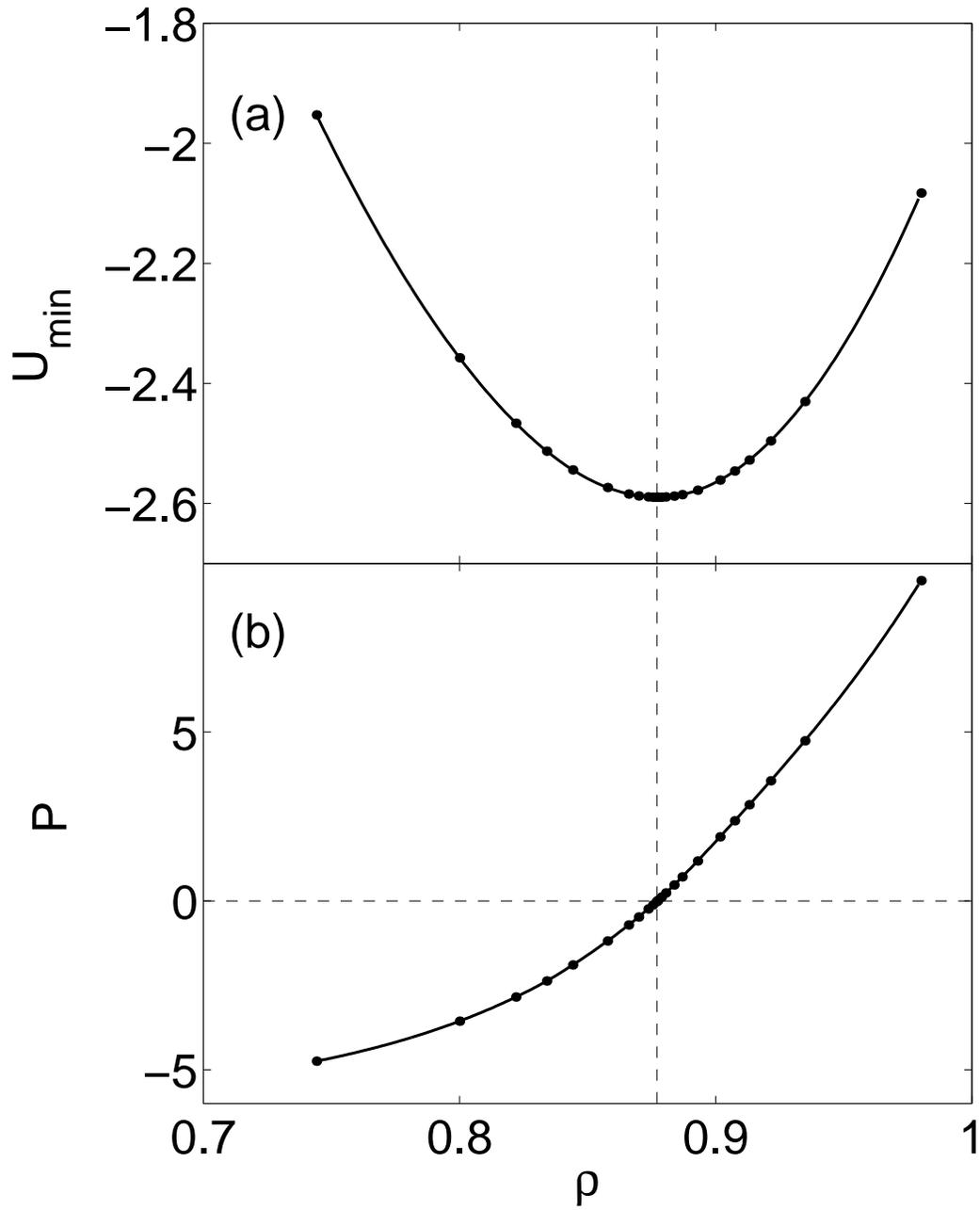,width=14cm,clip=}}
\vskip 20pt
\caption[]{(a) Minimal potential energy of the $\sigma$-phase
               structure as a function of density.  
           (b) Pressure as a function of density.  The dots show the
               data points and the solid curves are obtained by a cubic
               interpolation.}
\label{energy} 
\end{figure}

The atomic configuration of the $\sigma$ phase used in this study was
prepared as follows.  
A sample of $N=1620$ particles ($3\times3\times6$ unit cells) was
constructed by filling a tetragonal box of appropriate dimensions with
$\sigma$-phase unit cells.
We used the unit-cell atomic configuration suggested by G\"ahler
\cite{RothGaehler98,Gaehler} with $c/a=0.5176$.
The number density, $\rho=N/V$, $V=L_xL_yL_z$, of this atomic
configuration was $\rho=1.0048$.
This configuration was then used to provide the initial atomic
coordinates for variable-shape $NST$ (constant number of atoms,
pressure tensor, and temperature) molecular dynamics \cite{Smith96},
the run performed at $S=0$ and $T=0$.
This procedure is equivalent to a potential-energy minimization 
by the steepest descent method under the condition of independent 
pressure balance in each spatial dimension.  
The variable-shape $NST$ run resulted in an ideal crystalline
structure for which the fractional coordinates of the atoms in all
unit cells were identical within the precision of the calculations.
The structure of the $\sigma$ phase thus obtained is characterized by 
the minimum potential energy per atom,  $U_{{\rm min}}= -2.5899$, 
with respect to variations of thermodynamical parameters. 
In order to check this, we have performed similar $NST$ runs at different 
pressures and indeed found that the energy is minimum at zero pressure 
(see Fig.~\ref{energy}a,b). 
The density for the optimal structure has been found to be $\rho =
0.8771$ and the ratio of the lattice parameters is $c/a= 0.5273$,
close to that of $\beta$-U ($c/a = 0.5257$, at $T=720^{\circ}$C)
\cite{Thewlis52}.
The potential-energy minimum for the bcc structure at the same density
with the same potential was $U_{{\rm min}} = -2.6148$.
At zero pressure, the density of the bcc structure is $\rho= 0.8604$
and the minimum potential energy per atom is $U_{{\rm min}}
=-2.6357$, i.e. in both cases lower than for the $\sigma$ phase.
It was shown earlier \cite{Dzugutov97} that the potential energy per
atom for the $\sigma$ phase becomes lower than that for the bcc
structure at the same density as the temperature increases.
This is consistent with the fact that natural crystalline $\sigma$
phases are stable only at high temperatures.
They undergo a solid-solid phase transition to a simpler crystalline
phase as the temperature decreases.
In our simulations, however, the system was stable in the
range of temperatures $T \alt 0.9$ for as long as $t_{\rm run} = 5000$. 

We have also investigated the thermodynamical stability of the 
$\sigma$ phase under variable pressure. 
We have found that the $\sigma$ phase is stable for pressures 
in the range $-5 \alt P \alt 12$. 
At high pressures $P \agt 12$, a structural transformation occurs, 
resulting in the fcc structure. 
The phase diagram of the IC potential is not known at present.  We can
expect that at densities greater than the triple-point density for the
Lennard-Jones system, $\rho\approx0.85$ \cite{Hansen86}, the
solid-fluid coexistence curve for the IC system is close to that for
the LJ system.
We can only estimate that the melting temperature at $\rho=0.8771$ is
about $0.8 \alt T \alt 0.9$ from the fact that the $\sigma$-phase 
crystalline structure is stable at $T=0.8$ on the time scale of our
computations.
No diffusion was observed at temperatures up to $T=0.8$.  At $T=0.9$
the system stayed in a metastable superheated state for about 
$t_{\rm run}=5000$, after which it melted.

We used the density and the $c/a$ ratio obtained from the $NST$ run to
perform $NVE$ (constant number of atoms, volume, and total energy)
molecular dynamics runs starting from the three configurations 
mentioned above and scaling the velocities to zero at each time step,
which is also equivalent to a potential-energy minimization by the
steepest-descent method.
The same was done for one instantaneous configuration corresponding to
the temperature $T=0.8$.
The configurations resulting from this procedure were identical, which
is an indication that there is a well-defined potential-energy minimum
corresponding to a unique crystallographic arrangement of atoms within
the $\sigma$-phase unit cell.
This structure, scaled so that $a=c=1$, is shown in
fig.~\ref{sigmaMD}(b).
The atomic layers with $z$ close to 0.25 and 0.75, which are not
closely packed, show a small but significant puckering -- an effect
present in the $\beta$-U \cite{Lawson88} and Cr$_{48}$Fe$_{52}$
\cite{DaamsAtlas} structures.

\subsection{Vibrational dynamics}
\label{vibrations}

Above, we have discussed the similarities in the local structure of
the $\sigma$ phase and the IC glass. 
These similarities are expected to cause the vibrational spectra
in these two materials also to be similar. 
In order to check this assumption, in this subsection we 
investigate the vibrational properties of 
the $\sigma$ phase and compare the vibrational spectra of 
this crystal and the IC glass.

If the vibrational spectra are similar, the $\sigma$ phase can be
considered to be a good crystalline reference structure for the IC
glass.
One consequence of these similarities is that we can use the data about the
vibrational properticies of the $\sigma$ phase crystal to explain the
nature of the vibrational excitations in the corresponding amourphous
structure.
\subsubsection{Phonon dispersion in the $\sigma$ phase}
\label{dispersion}

\begin{figure}[b!]
\centerline{\epsfig{file=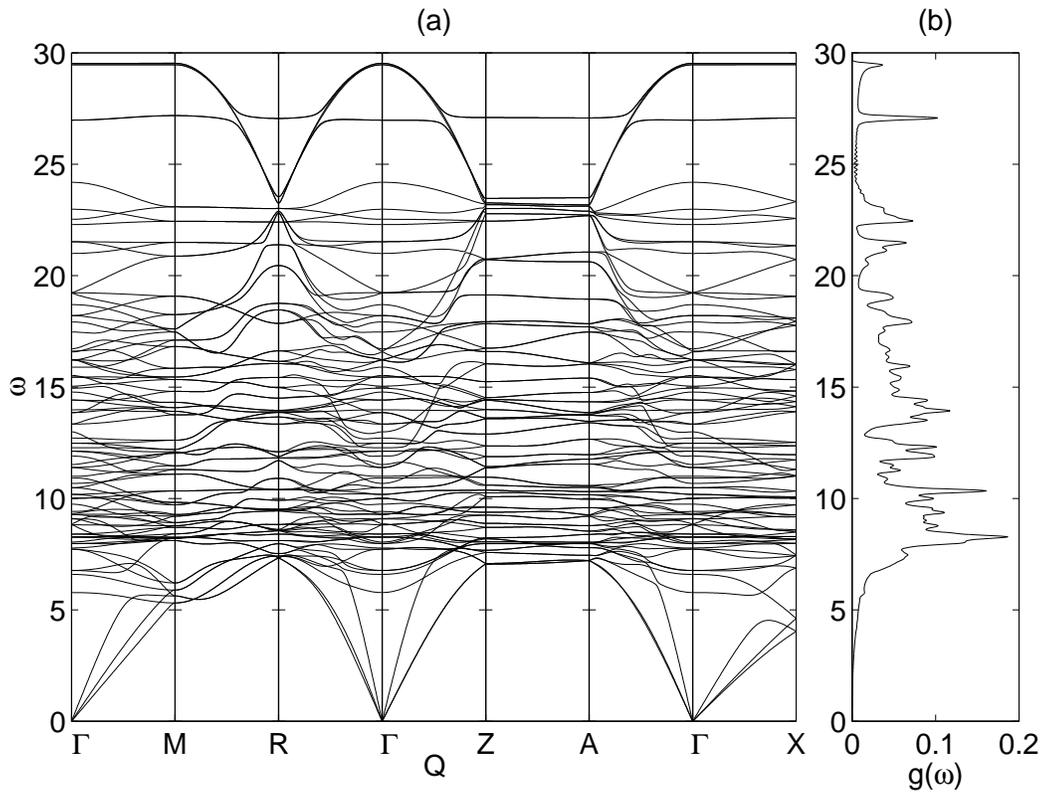,width=14cm,clip=}}
\vskip 20pt
\caption[]{Phonon-dispersion relations along the symmetry directions (a)
           and vibrational density of states (b) evaluated by a
           normal-mode analysis in the harmonic approximation. The
           symmetry points on the surface of the first Brillouin zone
           are $\Gamma=(0,0,0)$,
           ${\rm M}=(\frac{\pi}{a},\frac{\pi}{a},0)$,
           ${\rm R}=(\frac{\pi}{a},\frac{\pi}{a},\frac{\pi}{c})$,
           ${\rm Z}=(0,0,\frac{\pi}{c})$,
           ${\rm A}=(\frac{\pi}{a},0,\frac{\pi}{c})$,
           ${\rm X}=(\frac{\pi}{a},0,0)$.  }
\label{dispersionSigma} 
\end{figure}

The $\sigma$ phase has $30$ atoms per unit cell which result in $3$
acoustic and $87$ optic branches, as shown in
Fig.~\ref{dispersionSigma}.  
The vibrational density of states (VDOS) obtained by integration over
the first Brillouin zone (see Eq.~(\ref{vdos})) is shown on the right-hand
side of Fig.~\ref{dispersionSigma}.
The linear dispersion of the acoustic branches in the low-frequency range
($\omega \alt 4$) results in the Debye law for the VDOS, $g(\omega) =
3\omega^2/\omega_{\rm D}^3$, with the Debye frequency equal to
$\omega_{\rm D} \approx 23.89$.
The Debye frequency has been estimated from a fit of the initial part
of the VDOS by a parabolic function.  
The optic branches are densely distributed above the acoustic part.
There are no large gaps in the spectrum, which is a consequence of tight
binding and the mutual penetration of the basic structural units 
(Frank-Kasper polyhedra) in the $\sigma$ phase.
In other words, there are no isolated structural units, like molecules
in molecular crystals \cite{Dove93} and crystalline fullerens \cite{Yu93}, 
or tetrahedra in silica \cite{PRB2_97}, 
the vibrations of which form separate optic bands.
At some of the zone boundaries (e.g. the X point; see
Fig.~\ref{dispersionSigma}) the dispersion curves do not show zero
derivatives.
This is because the space group $P4_2/mnm$ of the $\sigma$ phase 
is nonsymmorphic, 
i.e. it contains nonpoint symmetry elements involving fractional
translations \cite{Dove93,Steurer93}. 
\subsubsection{Comparison with the IC glass}
\label{comparison}
\begin{figure}[b!]
\centerline{\epsfig{file=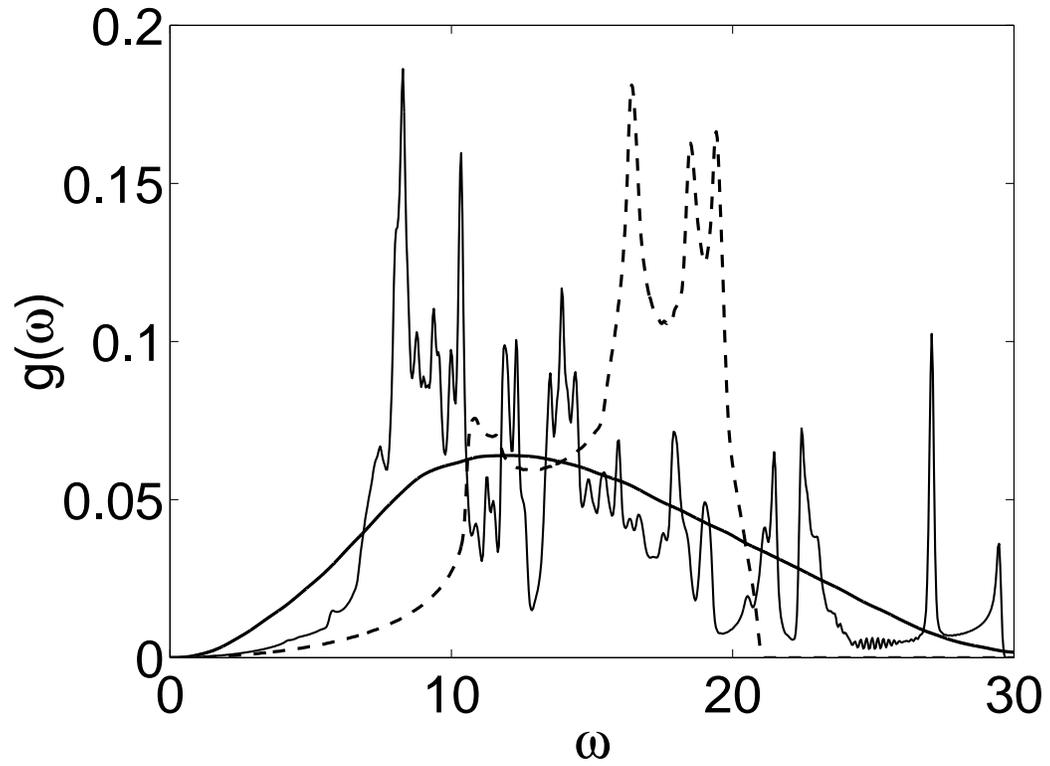,width=14cm,clip=}}
\vskip 20pt
\caption[]{Vibrational densities of states calculated by a normal-mode
           analysis for the $\sigma$-phase (thin solid line) and 
           for bcc (dashed line) structures, 
           and from the velocity autocorrelation
           function for the IC glass (thick solid line) 
           at the temperature $T=0.01$. 
          }
\label{VDOS_compare} 
\end{figure}

An informative characteristic of the vibrational dynamics in
the IC glass which can be compared with the $\sigma$ phase 
is its VDOS \cite{Dzugutov93_2} (see Fig.~\ref{VDOS_compare}) 
which can be easily obtained from the velocity autocorrelation 
function Eq.~(\ref{Z(t)}).
The VDOS for the bcc lattice is also presented for comparison
in the figure (the dashed line). 
We can clearly see that the frequency range 
of the whole spectrum is the same for the $\sigma$ phase 
and the IC glass but differs for the bcc lattice. 
The shape of the IC-glass spectrum mainly 
reproduces the basic features of the $\sigma$ phase 
spectrum and can be imagined as a 
superposition of broadened (by disorder) crystalline 
peaks. 
This is a consequence of the presence of a large number of optic 
modes in the vibrational spectrum of the sigma-phase structure 
located in the same frequency region as the whole spectrum of the IC glass. 
Therefore, the similarities in the VDOS of 
the $\sigma$ phase and the IC glass strongly support the 
choice of the $\sigma$ phase as a crystalline counterpart. 

\subsubsection{Soft modes in the $\sigma$ phase}
\label{subsubsec:softModes}
\begin{figure}[b!]
\centerline{\epsfig{file=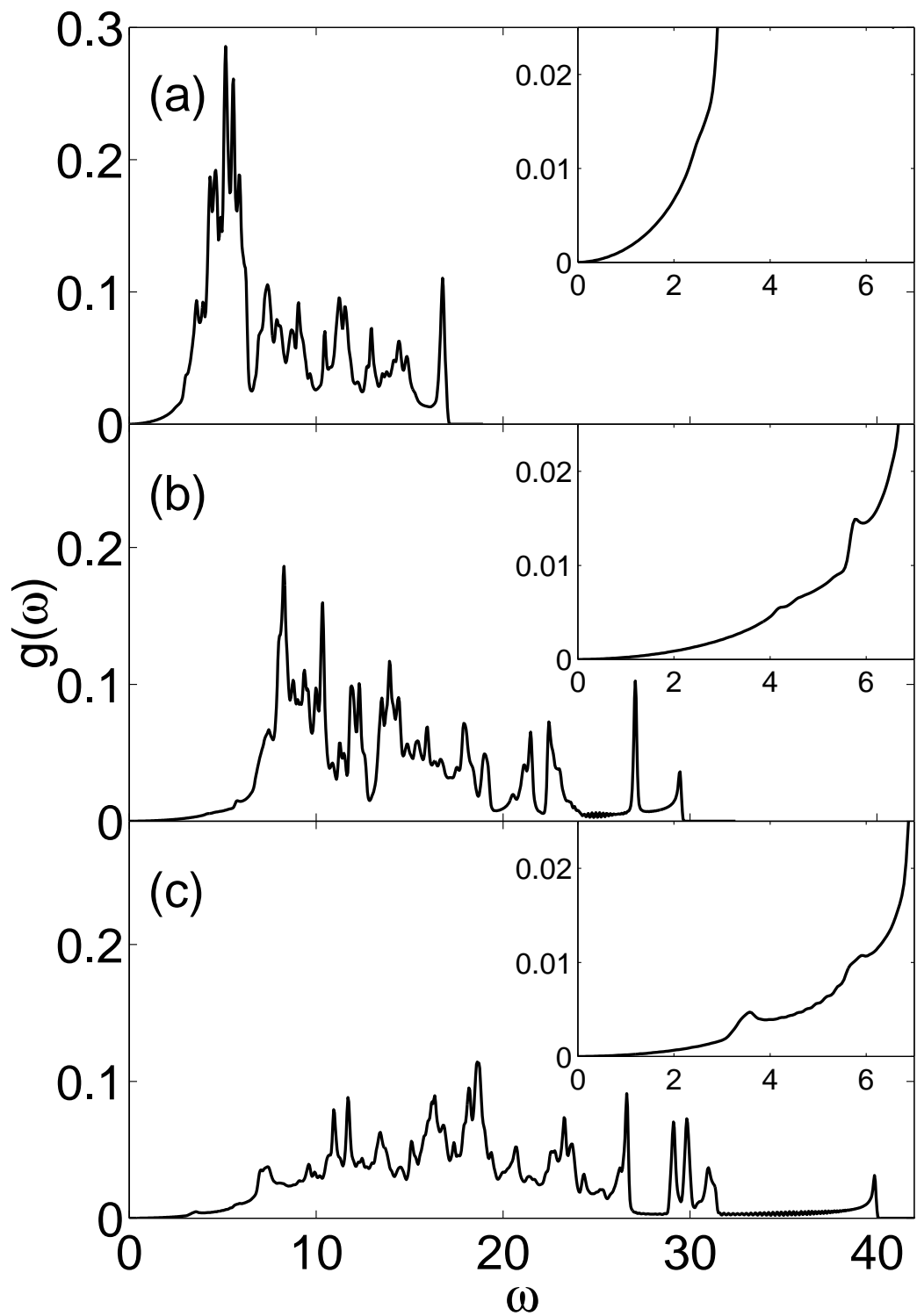,width=14cm,clip=}}
\vskip 20pt
\caption[]{Vibrational densities of states calculated by normal-mode
           analysis for the $\sigma$-phase structure at different 
           pressures: (a) $P = -4.74$, (b) $P = 0$, (c) $P = 9.49$.
           The insets show the low-frequency parts of the 
           corresponding spectra.
          }
\label{vdoses} 
\end{figure}

\begin{figure}[b!]
\centerline{\epsfig{file=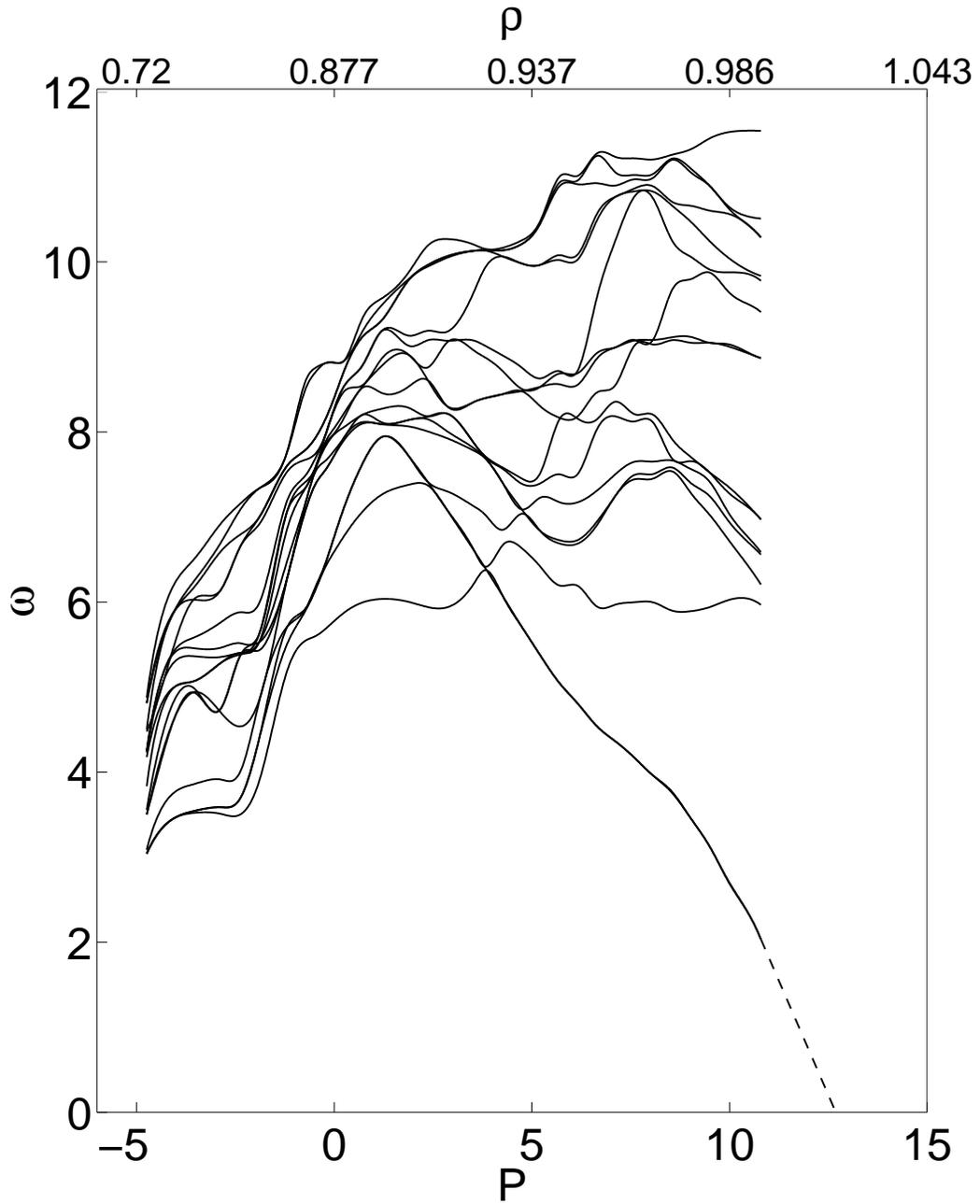,height=18cm,clip=}}
\vskip 20pt
\caption[]{Frequency of the low-frequency modes at the point $\Gamma$
           vs pressure.  The dashed line shows a linear
           extrapolation of the lowest-frequency curve.  }
\label{dispersionGamma} 
\end{figure}

As was mentioned in Sec.~\ref{S2b}, 
an interesting feature of atomic dynamics in 
the Frank-Kasper phases is related to the appearance of
low-frequency soft modes. 
We have investigated whether a soft mode 
appears in our model of the $\sigma$-phase structure.
For this purpose, we followed the evolution of the vibrational 
spectrum with variable pressure 
(see Fig.~\ref{vdoses} and, indeed, 
found that one of the lowest-frequency optic modes (doubly degenerate)
in the $\Gamma$-point shows soft-mode behavior. 
The frequency of this mode decreases both 
with decreasing and increasing pressure 
(see Fig.~\ref{dispersionGamma}). 
The decrease of the mode frequency at negative 
pressures is not surprising and reflects the 
softening of the whole vibrational spectrum 
(see Fig.~\ref{vdoses}a). 
However, with increasing pressure, the whole 
spectrum is shifted to higher frequencies 
(see Fig.~\ref{vdoses}c), 
while the frequency of the soft mode (a small peak 
around $\omega = 3.5$  in 
Fig.~\ref{vdoses}c) moves in the opposite 
direction, approaching zero and thus indicating 
a structural instability (structural phase transition to the fcc lattice)  
at a critical pressure $P_* \approx 12.5$ (see Fig.~\ref{dispersionGamma}). 
Around this value of the pressure the structure of the 
$\sigma$ phase becomes extremely unstable and an investigation of
the details of atomic motion requires a thorough analysis. 
We hope to address this point in another study. 
\subsubsection{Anharmonicity in the $\sigma$ phase}
\label{softmode}
One of the interesting questions concerning the vibrational dynamics
of the $\sigma$ phase is related to the range of applicability of
the harmonic approximation for the lattice vibrations.
We are able to anwer this question by investigating the vibrational
spectrum using the velocity autocorrelation function with increasing
temperature and comparing it with the results of the normal-mode
analysis (harmonic approximation).

\begin{figure}[b!]
\centerline{\epsfig{file=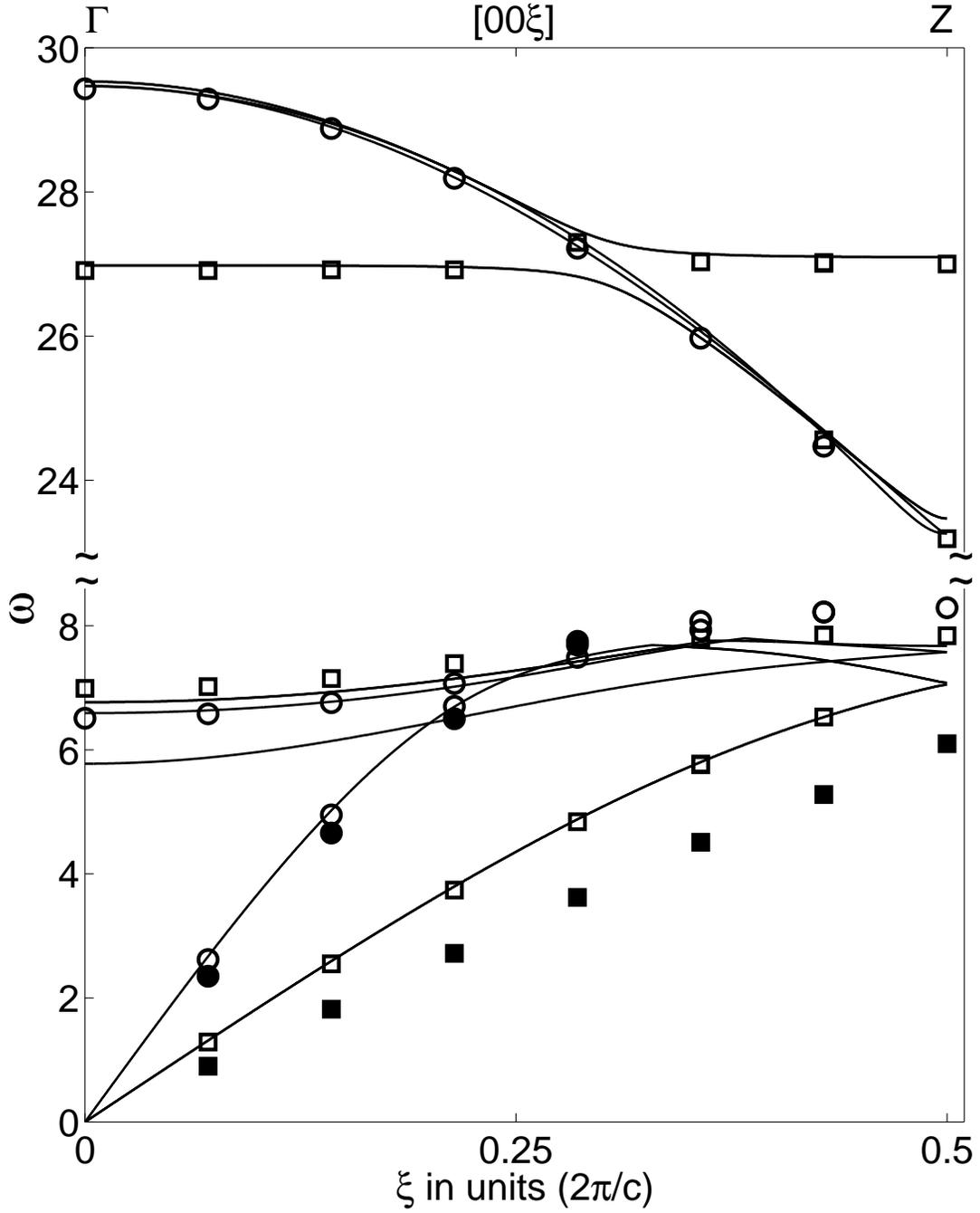,height=18cm,clip=}}
\vskip 20pt
\caption[]{Some phonon-dispersion curves calculated by a normal-mode
           analysis in the harmonic approximation (solid lines) and
           derived from the spectra of longitudinal and transverse
           particle-current autocorrelation functions (symbols) for the
           $\sigma$-phase structure. {\Large $\circ$}: longitudinal
           phonons; {\small $\square$}: transverse phonons at
           $T=0.01$. {\Large $\bullet$}: longitudinal phonons;
           $\blacksquare$: transverse phonons at $T=0.8$.  The linear
           size of the symbols is approximately equal to the width of the
           spectral peaks.  The direction of the wavevector
           ${\bf Q}=[0,0,\xi]$, $\Gamma$Z, is parallel to the
           $c$-axis. }
\label{dispersion_sigma} 
\end{figure}

To assess the degree of temperature-induced anharmonicity, we computed the 
dispersion relations for the symmetry direction
$[001]$ (${\bf Q} \parallel {\bf c}$, $\Gamma$Z in
Fig.~\ref{dispersionSigma}) at different temperatures by using 
both these techniques (molecular dynamics and normal-mode analysis) 
and compared the results.  
These are shown in Fig.~\ref{dispersion_sigma} for several  low-
and high-frequency dispersion branches 
for  two temperatures $T=0.01$ and $T=0.8$.  
At intermediate
frequencies, the density of dispersion branches is so high that a
comparison between the results of the two methods of calculation of
dispersion relations is hardly possible, mainly because of the finite
width of the respective peaks in the spectra of the current
autocorrelation functions (see sec.~\ref{sec:MD}).  Due to the fact
that the $\sigma$-phase space group $P4_2/mnm$ is nonsymmorphic,
i.e. it contains nonpoint symmetry elements involving fractional
translations \cite{Steurer93}, the phonon-dispersion relations,
derived from the peak positions in $C_l({\bf Q},\omega)$ and
$C_t({\bf Q},\omega)$, appear in the extended zone scheme
\cite{Dove93}. 
The optic modes cannot be measured in the vicinity of
the origin of the first Brillouin zone $Q=0$ ($Q=|{\bf Q}|$),
because this long-wavelength limit corresponds to motion of the system
as a whole, forbidden by the periodic boundary conditions.
Information about these modes is available at the boundaries
($Q=2\pi/c,6\pi/c$) and at the origin ($Q=4\pi/c$) of the second
extended Brillouin zone.  
The molecular dynamics results for the
dispersion relations at $T=0.01$ are adapted from the second extended
Brillouin zone.  To make possible the comparison with the 
results obtained in the harmonic
approximation, the data in the region $Q\in[5\pi/c,6\pi/c]$
were folded with respect to $Q=5\pi/c$ into the region
$Q\in[4\pi/c,5\pi/c]$, which corresponds to half of an irreducible
zone.  From these results we can see that the harmonic approximation
works quite well at the low temperature of $T=0.01$. At $T=0.8$, only
the acoustic branches could be resolved without ambiguity. 
Therefore, for this temperature, we used the data available from the first
Brillouin zone.  

\begin{figure}[b!]
\centerline{\epsfig{file=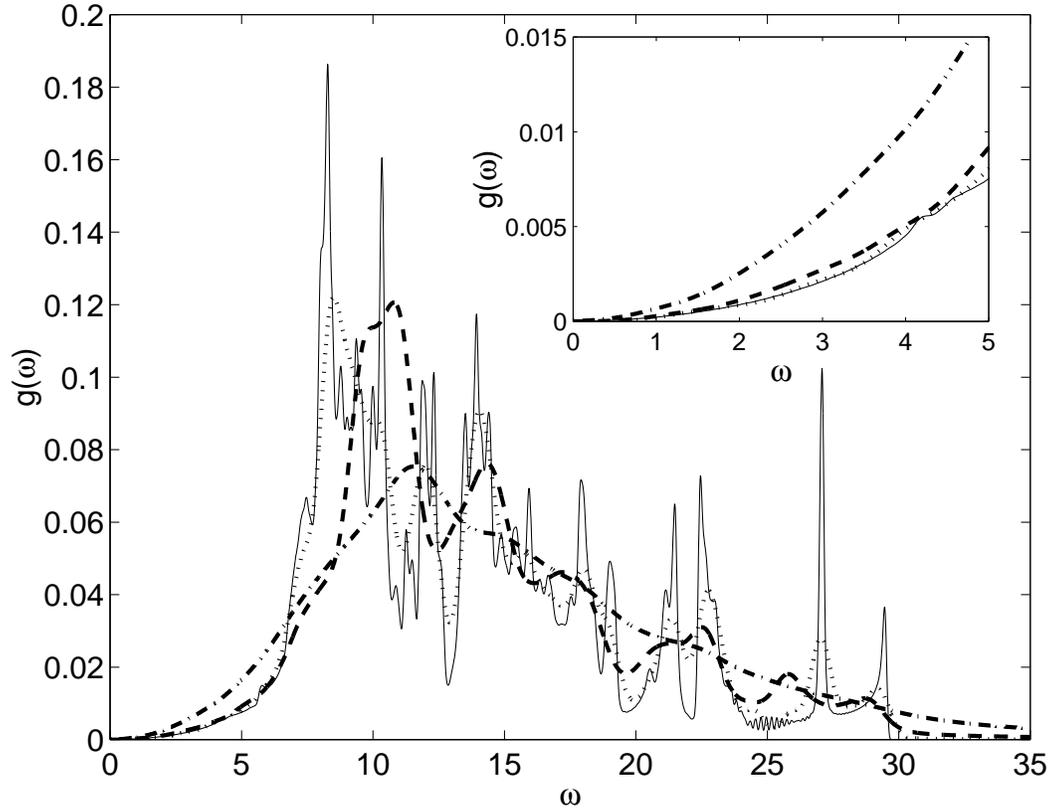,width=14cm,clip=}}
\vskip 20pt
\caption[]{Vibrational densities of states calculated by normal-mode
           analysis (solid line) and from the velocity autocorrelation
           function at different temperatures (broken lines) for the
           $\sigma$-phase structure. Dotted line: $T=0.01$, 
           dashed line: $T=0.2$, dashed-dotted line: $T=0.8$.
           The inset shows the low-frequency part of the spectrum}
\label{vdos} 
\end{figure}

One important signature of temperature-induced anharmonicity is a
softening of the acoustic modes, i.e. a lowering of the acoustic branches
with respect to those calculated in the harmonic approximation which
occurs as the temperature increases \cite{Dove93}.  
This effect can be clearly seen
for $T=0.8$ in Fig.~\ref{dispersion_sigma}.  
In accordance with this observation, 
the vibrational density of states for this temperature,
shown in Fig.~\ref{vdos}, exhibits the presence of excess states with
respect to the harmonic approximation.  
A deviation from the harmonic
approximation in $g(\omega)$ at low frequencies (see the inset in
Fig.~\ref{vdos}) starts to be noticeable at a temperature of about
$T=0.2$. 
Therefore, we can conclude the the lattice dynamics 
in the $\sigma$ phase is harmonic in a wide range of temperatures 
$T\alt 0.2$. 

\begin{figure}[b!]
\centerline{\epsfig{file=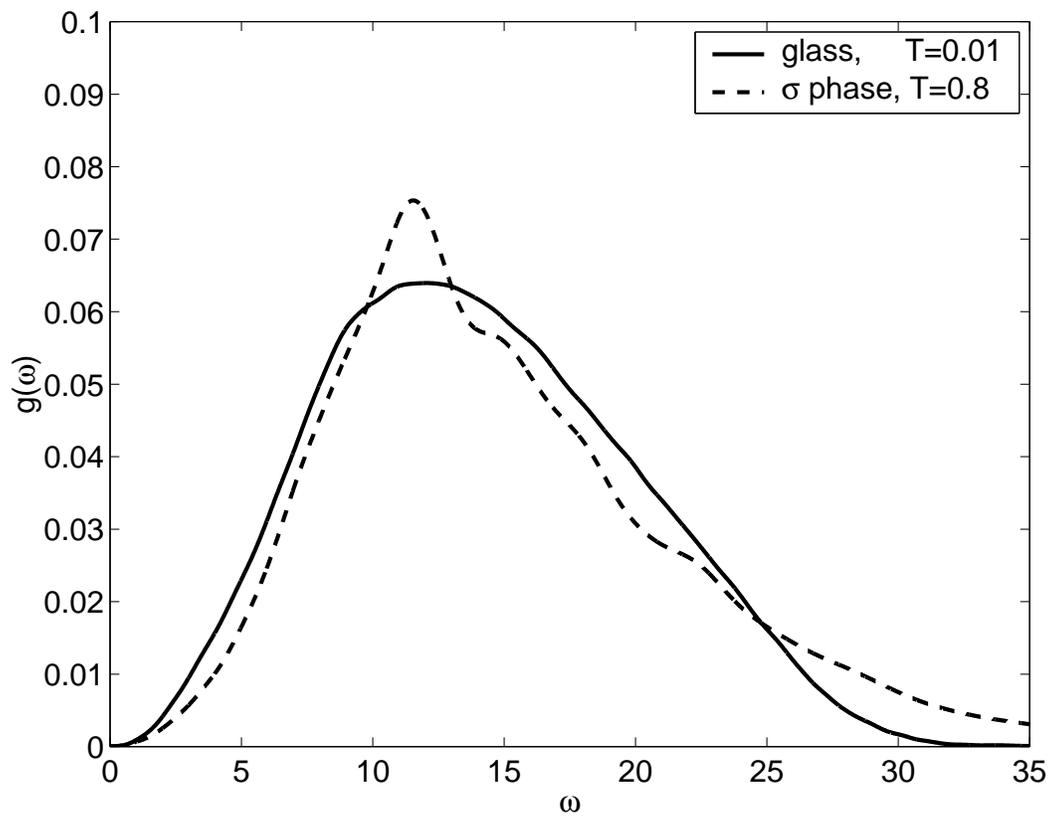,width=14cm,clip=}}
\vskip 20pt
\caption[]{Vibrational densities of states calculated from the velocity
           autocorrelation function for the $\sigma$-phase structure and
           the corresponding glass at the reduced temperatures shown.
           Both results are obtained at the same density
           $\rho=0.8771$.  }
\label{vdosGlass} 
\end{figure}

Finally, we would like to note the similarity between 
the high-temperature VDOS for 
the $\sigma$ phase and the low-temperature VDOS for the IC glass 
(see Fig.~\ref{vdosGlass}). 
Since $g(\omega)$ for the glass is only slightly
temperature dependent, we show it only for $T=0.01$.  
The fact that the densities of states for the glass and the high-temperature
$\sigma$ phase in fig.~\ref{vdosGlass} are remarkably similar clearly
indicates that the effect of the thermally-induced dynamical disorder
in the crystalline structure on the vibrational spectrum is similar to
that of the configurational disorder characteristic of the amorphous
structure.
\section{Conclusions}
\label{sec:conclusions}

In this paper we have studied the structural and vibrational properties of  
a $\sigma$-phase crystal. 
First, we have shown that it is possible to 
construct a structural model of a one-component 
$\sigma$ phase by means of molecular dynamics simulations 
using an appropriate pair potential. 
This $\sigma$-phase structure is stable in a wide range of 
thermodynamical parameters. 
Our model of the $\sigma$ phase contains only one atomic component. 
This is important in understanding the role of 
topological icosahedral order alone on the structural and 
dynamical properties and avoids the  effects arising from
the presence of different atomic species.  

Second, we have investigated atomic vibrational dynamics of the
$\sigma$ phase.
In particular, we have found the range of applicability of the harmonic
approximation in a description of atomic dynamics.
We have also demonstrated the existence of soft modes in the
$\sigma$ phase which leads to a structural phase transformation with
increasing pressure.

Third, we have demonstrated that the $\sigma$ phase is a good
crystalline counterpart of the IC glass.  This has been done on the
basis of a comparative analysis of the vibrational dynamics (vibrational
density of states).

We think that the results on the vibrational 
properties of the $\sigma$ phase discussed above can be 
used in an analysis of the peculiar vibrational properties 
of the IC glass (e.g. the Boson peak \cite{PRB2_97}). 
We also believe that the computational data of the vibrational
properties of the $\sigma$ phase could be of value for metallurgy
where this phase has received much detailed attention, chiefly because
of the detrimental effect which the formation of this phase has on
mechanical properties of certain steels \cite{PhysMet96}.

\section*{Acknowledgements}

S.I.S. and M.D. thank Trinity College for hospitality.
We are grateful to H.R.~Schober for bringing to our attention 
Ref.~\cite{Heiming91}.

\newpage
%
%

\newpage

%
%

\newpage

%
%

\newpage

%
%

\newpage

%
%

\newpage

%
%

\newpage

%
%

\newpage

%
%

\newpage

%
%

\newpage

%
%

\end{document}